\documentclass[aps, prd, reprint]{revtex4-2}   
\usepackage[utf8]{inputenc}
\usepackage[english]{babel}
\usepackage{amsmath,amssymb}
\usepackage{graphicx}
\usepackage{hyperref}
\usepackage{xcolor}
\usepackage{todonotes}
\usepackage{lipsum} 
\usepackage{booktabs}
\usepackage{multirow}
\usepackage{mathrsfs}
\usepackage{enumitem}
\usepackage{pifont}
\usepackage[cal=boondox, scr=rsfs]{mathalfa}

\begin{document}
\title{Massive Compact Stars Beyond the General Relativity Limit in Finslerian Gravity: A Possible Explanation for the GW190814 Secondary}
\author{Praveen J}
\email[Email:]{praveenjayarama1998@gmail.com}
\affiliation{Department of P.G. Studies and Research in Mathematics, Kuvempu University, Shanakaraghatta, Shivamogga,577451, India}
\affiliation{Pacif Institute of Cosmology and Selfology (PICS), Sagara, Sambalpur 768224, Odisha, INDIA}

\author{Rajesh Kumar}
\email[Email:]{rajesh.mathstat@ddugu.ac.in (Corresponding Author)}
\affiliation{Mathematical Astrophysics Lab\\
Department of Mathematics and Statistics, Deen Dayal Upadhyaya Gorakhpur University, U.P. 273009, India}

\author{Sadaf Fatima}
\email[Email:]{fsadaf00@gmail.com}
\affiliation{Mathematical Astrophysics Lab\\
Department of Mathematics and Statistics, Deen Dayal Upadhyaya Gorakhpur University, U.P. 273009, India}

\author{S K Narasimhamurthy}
\email[Email:]{sknmurthy@kuvempu.ac.in}
\affiliation{Department of P.G. Studies and Research in Mathematics, Kuvempu University, Shanakaraghatta, Shivamogga,577451, India}

\begin{abstract}
\textbf{Abstract:} 
The existence of an upper mass limit for compact stars is one of the fundamental predictions of general relativity, with important implications for the outcome of compact binary mergers and the nature of the proposed neutron star-black hole mass gap. The LIGO/Virgo collaboration announced the discovery of a compact binary merger, GW190814, containing a compact star with mass $2.5–2.67 M_\odot$ [R. Abbott et al.(2020) ApJ Lett., 896, L44], which provided an exciting new stimulus to the ongoing debate on whether/where a gap exists between the maximum mass of NS and the minimum mass of black hole. Such GW detection has also challenged conventional stellar models and renewed interest in exploring whether modified theories of gravity can accommodate such ultra-massive compact objects without invoking black hole formation. The present work investigated the structure and physical properties of compact stars within the framework of Finslerian gravity, employing the Heintzmann IIa gravitational potentials and showed that the our model can substantially enhance the maximum mass of compact stars upto $2.67\,M_{\odot}$, thereby offering a plausible explanation for massive compact objects such as the secondary component of GW190814. In addition, the moment of inertia is found to increase, indicating stronger rotational support and a redistribution of the internal mass profile. The analysis focuses on key astrophysical observables, including the mass-radius (M-R) and moment of inertia-mass (I-M) relations, and a detailed comparison with the general relativity (GR) counterpart is performed. Assuming a linear equation of state, we construct a class of physically viable anisotropic stellar models and analyze their behavior under the influence of Finslerian corrections. The physical viability of the model is rigorously tested through standard energy conditions, causality requirements, hydrostatic equilibrium governed by Tolman–Oppenheimer–Volkoff equation, and multiple stability diagnostics, including the adiabatic index, Harrison–Zeldovich–Novikov criterion, equation of state behavior, and cracking analysis.
\begin{description}
\item[Keywords]
Maximum mass, Neutron star, compact stars, mass-radius relation, moment of inertia, Finsler--Randers gravity, Heintzmann IIa potentials, anisotropy, modified gravity
\end{description}
\end{abstract}

\maketitle
\tableofcontents

\section{Introduction}\label{sec1}
Compact stars constitute natural laboratories for probing gravitational theories in regimes of extreme density and pressure, where the mutual interaction of gravity, matter, and spacetime geometry becomes highly nonlinear and nontrivial \cite{krori1975singularity, de1985collapse, sagert2006compact, maurya2015anisotropic}.  One of the most striking predictions of general relativity (GR) is the existence of a maximum mass for any static matter configuration. The maximum mass of nonrotating neutron stars is therefore a key quantity governing the outcome of compact binary mergers and provides stringent constraints on the poorly understood equation of state (EoS) of matter at supranuclear densities. The detection of the gravitational-wave (GW) event GW190814 by the LIGO/Virgo Collaboration, involving a compact object with mass $2.5–2.67 M_\odot$ ~\cite{abbott2020gw190814}, has generated significant interest in this context. This object lies within the putative mass gap between neutron stars and black holes, and its true nature remains uncertain. Several studies have explored this issue from different perspectives- Fattoyev et al.~\cite{fattoyev2020gw190814} argued that such a high mass is difficult to accommodate within conventional nucleonic EoSs, favoring a black hole interpretation. In contrast, Tan et al.~\cite{tan2020neutron} demonstrated that nontrivial behavior in the speed of sound, motivated by deconfined QCD matter, can support heavier neutron stars while remaining consistent with observational constraints. The observation of GW from an asymmetric binary opens the possibility for heavy neutron stars, but these pose challenges to models of the neutron star equation of state~\cite{most2020lower, fishbach2020does}. Other works have investigated the role of stiff EoSs, phase transitions, and phenomenological parameterizations in allowing neutron star such high masses~\cite{wu2020if, godzieba2021maximum}. Furthermore, rapidly rotating configurations have been proposed as a possible explanation for the secondary component of GW190814, which has been suggested that a superfast pulsar, rotating close to its Keplerian limit, could support masses in the range $2.5–2.67 M_\odot$ ~ \cite{zhang2020gw190814, breu2016maximum, wei2017rotating, most2020lower}. A stable neutron star with mass $2.6M_\odot$— and even higher—can be readily generated, challenge, however, is not to reconcile super-massive neutron stars,  but rather, to construct the stellar model and tidal deformability that favor suitable EoS~\cite{abbott2017gw170817, abbott2018macfoy}. 
\par
Despite various efforts, the existence and interpretation (stellar models) of such massive compact objects remain an open question, motivating the exploration of alternative gravitational frameworks and their implications for neutron star structure. Foundational investigations within GR established a robust theoretical framework through exact interior solutions and realistic equations of state, enabling detailed analyses of mass–radius bounds, stability criteria, and energy distributions \cite{herrera1992cracking, herrera1997local, knutsen1988stability, abreu2007sound, abubekerov2008mass, maurya2016new, maurya2015anisotropic, oertel2017equations, barbat2024comprehensive}. In recent years, significant attention has shifted toward modified theories of gravity and extended geometric frameworks—such as f(R), f(R,T), and f(Q) models—which aim to overcome certain limitations of standard general relativity and provide alternative explanations for observed high-mass compact objects \cite{kumar2024anisotropic, sharif2023compact, sharif2024impact, kumar2022relativistic, kumar2024anisotropic, gul2024viable, gul2024viablea, rej2024well, rej2021charged, kumar2024anisotropic1, kumar2025constructing}. Collectively, these developments reflect the progressive refinement of compact star modeling and emphasize their crucial role in connecting theoretical predictions with astrophysical observations.
\par
Among the various alternative frameworks, Finslerian gravity emerges as a natural geometric extension of GR, grounded in Finsler geometry where the spacetime metric depends not only on position but also on the direction (or velocity) of particles \cite{antonelli2003handbook, bao2012introduction, bogoslovsky1999finslerian, vacaru1997superstrings}. The development of relativistic theories in Finsler geometry has demonstrated that spacetime structures can be described by Bogoslovsky-type metrics, leading to modified field equations and inherently anisotropic gravitational features \cite{kouretsis2009general, basilakos2013cosmological, basilakos2013resembling, pfeifer2012finsler, pfeifer2011causal, kouretsis2010imperfect, kouretsis2012covariant, stavrinos2002remarks, stavrinos2008friedman, stavrinos2012weak}. A particularly important subclass is the \textit{Finsler–Randers (FR)} metric, originally introduced by Randers \cite{randers1941asymmetrical}, which provides a direct extension of Riemannian geometry through the inclusion of a linear one-form. In this formulation, the metric function takes the form $F(x,y)=\alpha(x,y)+\beta(x,y)$, where $\alpha(x,y)$ represents the Riemannian component and $\beta(x,y)$ is a linear term. The presence of $\beta$ induces intrinsic anisotropy and directional dependence in the spacetime structure, making the FR geometry particularly well-suited for modeling anisotropic physical systems~\cite{basilakos2013cosmological, papagiannopoulos2017finsler, triantafyllopoulos2020schwarzschild}. Within the framework of Finslerian gravity (FR gravity), compact objects have been widely investigated, revealing that intrinsic geometric anisotropy induces notable modifications in internal structure, equilibrium conditions, horizon structure, thermodynamic behavior, and traversability criteria, often alleviating the necessity for exotic matter \cite{rahaman2012strange, rahaman2016finslerian, das2023possible, vacaru2010finsler, narasimhamurthy2026influence, nekouee2024black, praveen2024exploring, rahaman2015finslerian}.

\par
The organization of the manuscript is as follows. Section~\ref{sec2} introduces the fundamental geometric framework of Finsler--Randers gravity, and the corresponding modified gravitational field equations are presented. In Section~\ref{sec3}, compact stellar models are formulated within this framework by employing a linear equation of state along with the Heintzmann IIa gravitational potentials. Section~\ref{sec7}, discussed the prediction of maximum mass, compactness and moment of inertia through the $M-R$ and $I-M$ relations and compared with the corresponding GR results to highlight the observation implications of the model. Section~\ref{sec4} focuses on the implementation of boundary conditions and central regularity requirements to construct the compact stars' model in FR gravity. The physical behavior of the stellar configuration is analyzed in Section~\ref{sec5}, including energy density, pressures, anisotropy, energy conditions, gravitational mass, compactness and red-shift. Section~\ref{sec6} is  examines the equilibrium and stability of the system through the energy conditions, EoS parameter, causality condition, Tolman--Oppenheimer--Volkoff equation,  adiabatic index, and Harrison--Zeldovich--Novikov criterion.  Finally, Section~\ref{sec8} summarizes the conclusions and key outcomes of the study.

\section{Mathematical formalism of Finsler--Randers Geometry}\label{sec2}
Finsler geometry constitutes a intrinsic generalization of Riemannian geometry by permitting the spacetime metric to depend not only on the positional coordinates $x^{\mu}$ but also on the directional components $y^{\mu}$ in the tangent bundle $TM$. In this formalism, the spacetime interval is governed by a positively homogeneous function $F(x,y)$ defined on $TM$, thereby introducing an intrinsic dependence on both position and direction~\cite{bao2012introduction,rund2012differential}
\begin{equation}
ds = F(x,y), \qquad y^{\mu} = \frac{dx^{\mu}}{d\tau},
\end{equation}
which is smooth on the slit tangent bundle $TM\setminus\{0\}$ and satisfies the homogeneity condition $F(x,\lambda y)=\lambda F(x,y)$ for $\lambda>0$.

The corresponding Finsler metric is defined through the Hessian of $F^2$ with respect to the directional variables,
\begin{equation}
g^{F}_{\mu\nu}(x,y) = \frac{1}{2}\frac{\partial^2 F^2}{\partial y^{\mu}\partial y^{\nu}},
\end{equation}
thereby generating a symmetric non–degenerate tensor field on the tangent bundle. In contrast to Riemannian geometry, the metric tensor therefore acquires an intrinsic directional dependence, reflecting a local anisotropic structure of spacetime.

Among the various realizations of Finsler geometry, Randers spaces constitute a particularly significant subclass that introduces anisotropy through a vector field contribution \cite{randers1941asymmetrical}.
\par
The Finslerian structure function can be described as \cite{matsumoto1992theory}
\begin{equation}\label{eq3}
F(x,y)=\alpha(x,y)+\beta(x,y),
\end{equation}
where
\begin{equation}
\alpha(x,y)=\sqrt{g_{\mu\nu}(x)y^{\mu}y^{\nu}}, \qquad
\beta(x,y)=b_{\mu}(x)y^{\mu}.
\end{equation}
Here $g_{\mu\nu}$ denotes the underlying Riemannian metric, while the vector field $b_{\mu}$ describes the deviation from isotropy.

For a weak anisotropic background described by a small time–like vector field, this vector field becomes \cite{stavrinos2008friedman, stavrinos2002remarks, kouretsis2009general, kouretsis2010imperfect, praveen2025cosmological}
\begin{equation}
b_{\mu}=(\xi(r),0,0,0), \qquad |\xi|\ll1 ,
\end{equation}
which introduces a perturbative Finslerian correction to the spacetime geometry. In order to perform computation within a Riemannian framework, we adopt the osculating Riemannian approach \cite{rund2012differential, asanov2012finsler}, whereby the direction dependence is evaluated along a smooth vector field $y^{\mu}=U^{\mu}(x)$. The effective Finslerian spacetime metric is then given by
\begin{equation}
\tilde g_{\mu\nu}(x)=g^{F}_{\mu\nu}\big(x,U(x)\big).
\end{equation}
Under the weak–anisotropy approximation the Finsler--Randers (FR) metric components becomes 
\begin{align}
\tilde g_{00} &= g_{00}+2\sqrt{g_{00}}\xi+\xi^2, \label{eq7}\\
\tilde g_{ii} &= g_{ii}\!\left(1+\frac{\xi}{\sqrt{g_{00}}}\right), \qquad i=1,2,3. \label{eq8}
\end{align}
This construction effectively embeds the Finslerian anisotropic corrections into an equivalent Riemannian spacetime, enabling the use of standard gravitational tools while preserving the directional imprint of the underlying geometry.
\par
The corresponding affine connection follows from the Levi–Civita construction becomes
\begin{equation}
\Gamma^{\lambda}_{\mu\nu}=\frac{1}{2}\tilde g^{\lambda\sigma}\left(\partial_{\mu}\tilde g_{\sigma\nu}+\partial_{\nu}\tilde g_{\sigma\mu}-\partial_{\sigma}\tilde g_{\mu\nu}\right),
\end{equation}
from which the curvature tensor, Ricci tensor and scalar curvature are obtained through the usual geometric definitions, 
\begin{align}
   \mathbb{R}^{\lambda}{}_{\mu \nu \delta} &= \partial_{\nu}\Gamma^{\lambda}{}_{\mu \delta} - \partial_{\delta}\Gamma^{\lambda}{}_{\mu \nu}
   + \Gamma^{m}{}_{\mu \delta}\Gamma^{\lambda}{}_{m \nu} - \Gamma^{m}{}_{\mu \nu }\Gamma^{\lambda}{}_{m \delta}, \\[3pt]
   \mathbb{R}_{\mu \nu } &= \mathbb{R}^{\lambda}{}_{\mu \lambda \nu}, 
   \qquad 
   \mathbb{R} = \tilde g^{\mu \nu} \mathbb{R}_{\mu \nu}.
\end{align}

where $\lambda, \mu, \nu, m = 0,1,2,3$. In such manner, the gravitational dynamics inherit small but physically meaningful anisotropic corrections arising from the FR structure of spacetime.
\section{Compact Stellar Configurations in Finsler–Randers framework}\label{sec3}
To describe static compact stellar configurations we consider the general spherically symmetric line element \cite{adler1974fluid, kumar2025anisotropic, kumar2024anisotropic}
\begin{equation}
ds^{2}_{-}=g_{ij} dx^i dx^j=e^{\mu(r)}dt^{2}-e^{\nu(r)}dr^{2}
-r^{2}(d\theta^{2}+\sin^{2}\theta\,d\phi^{2})
\label{eq16a}
\end{equation}
where the metric potentials $\mu(r)$ and $\nu(r)$ characterize the gravitational potentials inside the stellar interior, which constitutes the Riemannian metric \(\alpha\) part of the FR metric~\eqref{eq3}.

\par
Upon embedding the Finsler–Randers correction within the osculating framework, the effective spacetime metric components presented in Eqs.~\eqref{eq7}–\eqref{eq8} take the following form,
\begin{align}
\tilde g_{00} &= e^{\mu(r)}+2e^{\mu(r)/2}\xi+\xi^2, \label{eq11}\\
\tilde g_{11} &= e^{\nu(r)}\!\left(1+\frac{\xi}{e^{\mu(r)/2}}\right), \label{eq12}\\
\tilde g_{22} &= r^2\!\left(1+\frac{\xi}{e^{\mu(r)/2}}\right), \label{eq13}\\
\tilde g_{33} &= r^2\sin^2\theta\left(1+\frac{\xi}{e^{\mu(r)/2}}\right). \label{eq14}
\end{align}

The gravitational field equation is, 
\begin{equation}\label{eq15}
R_{\mu\nu}=8\pi\left(T_{\mu\nu}-\frac12 T\,\tilde g_{\mu\nu}\right),
\end{equation}
(assuming $G=1=c$) and  $T_{\mu\nu}$ is the interior matter distribution is described by an anisotropic fluid energy–momentum tensor
\begin{equation}
T_{\mu\nu}=(\rho+p_t)U_{\mu}U_{\nu}-p_t\,\tilde g_{\mu\nu}+(p_r-_t) \chi_{\mu} \chi_{\nu}.
\end{equation}
Here $\rho$ represents the energy density, while $p_r$ and $p_t$ denote the radial and tangential pressures, respectively. The vectors $U^{\mu}$ and $\chi^{\mu}$ correspond to the four–velocity of the fluid and the unit radial vector satisfying $U^{\mu}U_{\mu}=1$ and $\chi^{\mu}\chi_{\mu}=-1$.

\par
In view of eqs.~\eqref{eq11}-\eqref{eq14}, the field equation~\eqref{eq15} yields the following
\begin{widetext}
\begin{align}
    8 \pi \rho &= \frac{1 - e^{-\nu} + e^{-\nu}r\nu' + e^{-\nu - \frac{\mu}{2}}r\!\left(-2 + \tfrac{1}{2}r\nu' + r\mu'\right)\xi'}{r^2}, \label{eq17} \\
   8 \pi p_r &= \frac{-2 + 2e^{-\nu} + 2e^{-\nu}r\mu' + e^{-\nu - \frac{\mu}{2}}r(3 + r\mu')\xi'}{2r^2},  \label{eq18}\\
   8\pi p_t &= \frac{e^{-\nu - \frac{\mu}{2}}(6 - 3r\nu' - 4r\mu')\xi' + 2e^{-\nu}\!\left[(-\nu' + \mu')(2 + r\mu') + 2r\mu''\right]}{8r}.  \label{eq19}
\end{align}
\end{widetext}
Since the field equations~(\ref{eq17})–(\ref{eq19}) contain six unknown functions, namely $(\rho,p_r,p_t,\mu,\nu,\xi)$, the system is inherently underdetermined. To render the system solvable, it is necessary to introduce three additional physically motivated constraints. In order to do this, we adopt a \textit{linear equation of state} in conjunction with the \textit{Heintzmann IIa metric potentials} as auxiliary conditions, which enable us to obtain a consistent and closed-form solution of the field equations.

\subsection{Linear equation of state}\label{sec3.1}
In the absence of a well-established microphysical equation of state (EoS) for ultra-dense matter, phenomenological relations between pressure and energy density are commonly employed to model the internal structure of compact stars. Among these, the linear EoS
\begin{equation}\label{eq20}
p_r = a\,\rho - b,
\end{equation}
where $a$ and $b$ are constants, has been widely utilized owing to its analytical simplicity and its effectiveness in capturing the essential physics of matter at extreme densities. The dimensionless parameter $a$ characterizes the stiffness of the matter distribution, while $b$ (expressed in units of $ km^{-2}$ or $MeV/fm^3$ can be interpreted as a vacuum pressure or bag constant that shifts the pressure–density relation. Such a form of EoS naturally emerges as an effective description of relativistic fluids, quark matter, and self-bound stellar configurations, and has been shown to yield realistic mass–radius relations consistent with astrophysical observations~\cite{mak2002exact, ivanov2017analytical}.
\par
Within the framework of FR gravity, where geometric anisotropy inherently induces pressure anisotropy, the adoption of a linear EoS offers a systematic and physically transparent approach to investigate the influence of spacetime anisotropy on the internal structure and stability of compact objects. Consequently, it serves as a well-motivated and widely accepted closure condition for constructing realistic and observationally consistent stellar models in both general relativity and modified gravity scenarios~\cite{thirukkanesh2008charged, sharma2007class}. Moreover, in anisotropic configurations, this form of EoS proves particularly advantageous, as it facilitates the closure of the highly nonlinear field equations without the need for arbitrary assumptions regarding the anisotropy profile.

Substituting Eqs.~\eqref{eq17}–\eqref{eq18} into Eq.~\eqref{eq20}, the resulting system enables the determination of the anisotropic function $\xi(r)$, as follows
\begin{equation}\label{eq21}
\xi'(r) =-\frac{2 e^{\mu/2}\!\left[-1 + e^{\nu} - a + e^{\nu}s - e^{\nu}r^{2}b + ra\,\nu' - r\,\mu'\right]}{r\!\left[-3 - 4a + ra\,\nu' - r\,\mu' + 2ra\,\mu'\right]} .
\end{equation}
\subsection{Heintzmann IIa gravitational potentials}\label{sec3.2}
The Heintzmann IIa gravitational potentials, introduced by Hermann Heintzmann, represent an important class of exact interior solutions to the  field equations for static, spherically symmetric compact objects~\cite{heintzmann1975neutron, heintzmann1969new, delgaty1998physical}. In this formulation, the temporal and radial metric potential is prescribed in a well-behaved analytic form that remains finite and regular at the stellar center. Owing to these appealing features, the Heintzmann IIa solution has been widely employed as a viable model for describing neutron stars and other ultra-dense compact objects, and has further been generalized in studies involving anisotropic matter distributions and modified theories of gravity. The Heintzmann IIa metric potentials are given by 
\begin{equation}\label{eq22}
e^{\mu(r)} = A^{2}(1 + Br^{2})^{3},
\end{equation}
\begin{equation}\label{eq23}
e^{\nu(r)} = \frac{1}{1 - \frac{3Br^{2}}{2}
\left(\frac{1 + \frac{F}{\sqrt{1 + 4Br^{2}}}}{1 + Br^{2}}\right)},
\end{equation}
where \(A\) and \(F\) are dimensionless constants, while \(B\) carries the dimension of $km^{-2}$. This metric potentials have been extensively employed in the modeling of compact stellar interiors due to its mathematical tractability and its ability to generate physically acceptable solutions of the Einstein field equations~\cite{waseem2024isotropic, rej2024well}.

\par
Now using the above  Eqs. \eqref{eq22}-\eqref{eq23} into Eq.~\eqref{eq20}, one can obtain
\begin{equation}
\xi'(r)=\frac{Q_1(r)}{Q_2(r)}
\label{eq23a}
\end{equation}
where
\begin{widetext}
\begin{align}
Q_1(r) &= 2 r \sqrt{A^2 (B r^2+1)^3}\Big[4 B^3 r^4 \big(\sqrt{4 B r^2+1}(3a+2b r^2+9)+21F\big)+3 B^2 r^2 \big(\sqrt{4 B r^2+1}(13a+6b r^2-9)+(9a+11)F\big)  \notag\\
&\quad +3 B \big(\sqrt{4 B r^2+1}(3a+4b r^2-3)+3aF+F\big)
+2b\sqrt{4 B r^2+1}
\Big],
\end{align}
\begin{align}
Q_2(r) &= 4 B^3 r^6 \big((8a-9)\sqrt{4 B r^2+1}+27(a-1)F\big) -3 B^2 r^4 \big((32a-17)\sqrt{4 B r^2+1}+3(4a+7)F\big) \notag\\
&\quad +3 B r^2 \big((2a+13)\sqrt{4 B r^2+1}-3(2a+1)F\big)+2(4a+3)\sqrt{4 B r^2+1}.
\end{align}
\end{widetext}

Now making use Eqs.~\eqref{eq22}-\eqref{eq23} along with Eq. (\ref{eq23a}) into field equations \eqref{eq17})–\eqref{eq18}, one obtain
\begin{widetext}
\begin{align}
8\pi \rho = &\Bigg[-16 b-16 B^6 r^{10} (16 b r^2+99)-12 B^5 r^8 \Big(2 F S(r) (9 b r^2+82)+189 F^2-2 (8 b r^2+147)\Big) \notag\\
&-3 B^4 r^6 \Big(4 F S(r) (30 b r^2-151)+99 F^2 -432 b r^2
+129\Big) +2 B^3 r^4 \Big(-9 F S(r) (2 b r^2+15)+458 b r^2+531\Big) \notag\\
&+3 B^2 r^2 \Big(4 F S(r) (12 b r^2-11)-9 F^2-8 b r^2
+285\Big)+6 B \Big(F S(r) (6 b r^2+5)-18 b r^2+21\Big) \Bigg] 
\Bigg/ \notag\\ 
& \Bigg[2 (B r^2+1)^2\Big(8 a+B r^2 \big(38 a+16 (8 a-9) B^3 r^6   +4 B^2 r^4 (-88 a+27 (a-1) F S(r)+42) \notag\\
&\qquad -9 B r^2 (8 a +(4 a+7) F S(r)-23)  \notag\\
&\qquad -9 (2 a+1) F S(r)+63\big)+6\Big)\Bigg].
\end{align}

\begin{align}
8\pi p_r =&\Bigg[\left(6Br^2/(Br^2+1)+3\right)\left(1-\left(3Br^2(F/S(r)+1)\right)/(2(Br^2+1))\right) \Big(
4B^3r^4\left(S(r)(3a+2b r^2+9)+21F\right) \nonumber\\
&\quad +3B^2r^2\left(S(r)(13a+6b r^2-9)+(9a+11)F\right) +3B\left(S(r)(3a+4b r^2-3)+3a F+F\right) \nonumber\\
&\quad +2b S(r)\Big)\Bigg]\Bigg/ \nonumber\\
& 
\Bigg[4B^3r^6\left((8a-9)S(r)+27(a-1)F\right) -3B^2r^4\left((32a-17)S(r)+3(4a+7)F\right) +3Br^2\left((2a+13)S(r)-3(2a+1)F\right) \nonumber\\
&\quad +2(4a+3)S(r)\Bigg]  + \frac{3B\left(Br^2(-3F/S(r)-1)+2\right)}{(Br^2+1)^2}  -\frac{3B\left(F/S(r)+1\right)}{(2(Br^2+1))}
\end{align}

\begin{align}
p_{t}=&3\Bigg[4b + B\Big(66a+192B^6r^{12}(-13a+2b r^2+27) +8B^5r^{10}\Big(F S(r)(-485a+42b r^2+891) \nonumber\\
& +63(19-12a)F^2-2(-687a+14b r^2+801)\Big) \nonumber\\
&+2B^4r^8\Big(F S(r)(4667a+342b r^2-3333)+9(43a+363)F^2-18(257a+62b r^2+33)\Big) \nonumber\\
&+B^3r^6\Big(-7227a+2F S(r)(604a+147b r^2-3252)+9(191a+159)F^2
-2298b r^2+5877\Big) \nonumber\\
&-B^2r^4\Big(87a+2F S(r)(815a+66b r^2+801)-9(25a+11)F^2
+688b r^2-2547\Big) \nonumber\\
&-2Br^2\Big(F S(r)(157a+45b r^2+75)+9a F^2-9(28a+b r^2+21)
\Big) \nonumber\\
&+2F S(r)(a-6b r^2-3)+36b r^2+18\Big)\Bigg] \Bigg/ \nonumber\\
&\Bigg[4(Br^2+1)^2(4Br^2+1)\Big(8a+Br^2\Big(38a+16(8a-9)B^3r^6 +\quad +4B^2r^4(-88a+27(a-1)F S(r)+42) \nonumber\\
&\quad -9Br^2(8a+(4a+7)F S(r)-23)  -9(2a+1)F S(r)+63\Big)+6\Bigg].
\end{align}
\end{widetext}

where \(S(r)=\sqrt{4Br^2+1}\). 
\par
To construct a physically consistent compact star model, the constants \(A\), \(B\), \(F\), \(a\) and \(b\) are fixed by imposing appropriate boundary conditions, (obtained by matching the interior solution to the exterior Schwarzschild spacetime at the stellar surface $r=R$) and central isotropy condition ar $r=0$ (see Sec.~\ref{sec4} onward). 

\section{$M-R$ and $I-M$ Relationship: Prediction of maximum mass ($2.67 M_\odot$) and moment of inertia ($2.71 \times 10^{45}\,\mathrm{g\,cm^2}$)}\label{sec7}
The structural and rotational properties of compact stars can be systematically analyzed through the mass–radius $M-R$ and moment of inertia–mass $I-M$ relations. Observational constraints on these correlations serve as a vital link between theoretical models and astrophysical measurements, enabling direct tests of compact star configurations by connecting their internal structure to observable quantities~\cite{lattimer2004physics}. By confronting the predicted $M-R$ and $I-M$ curves with current observational bounds, the consistency of the model can be assessed and identifying potential geometric signatures beyond standard GR. Such predictions not only establish the physical viability of the proposed configurations but also elucidate the role of anisotropy in influencing the observable characteristics of compact stars.
\subsection{$M-R$ Relation: A Comparative analysis with GR and its Astrophysical Relevance}
The mass–radius (M–R) relation serves as one of the most informative diagnostics for probing the equilibrium structure of compact stars corresponding to a given EoS. By relating the total gravitational mass to the associated stellar radius obtained from the interior field equations, it provides a direct measure of the capability of a gravitational framework to support ultra-dense matter against collapse. Since the interior densities of neutron stars far exceed those achievable in terrestrial experiments, theoretical $M–R$ relations play a crucial role in understanding matter under extreme conditions and in assessing the consistency of a given model with astrophysical observations.
\par
\begin{figure*}[]
\centering
\setlength{\tabcolsep}{6pt}
\renewcommand{\arraystretch}{1.0}
\begin{tabular}{cc}
\includegraphics[width=0.48\textwidth]{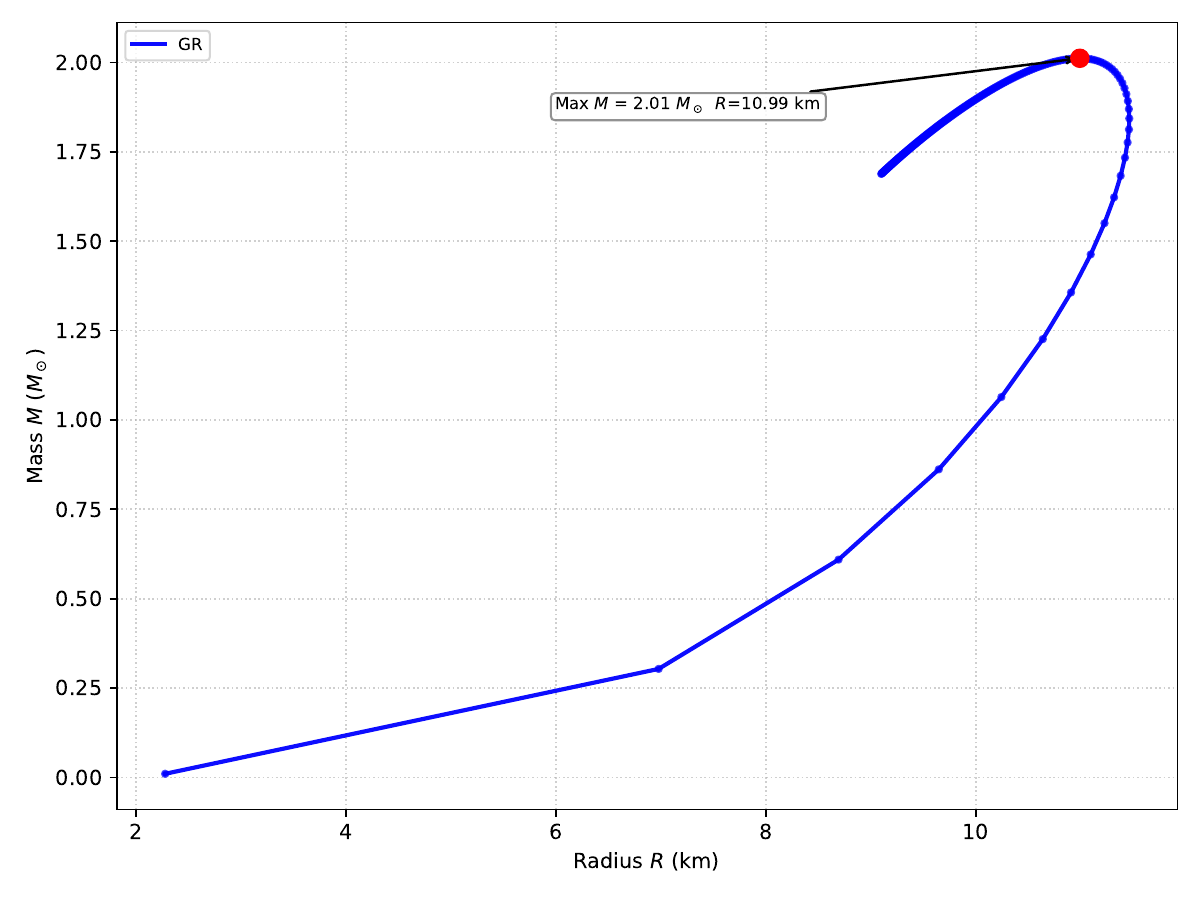} &
\includegraphics[width=0.48\textwidth]{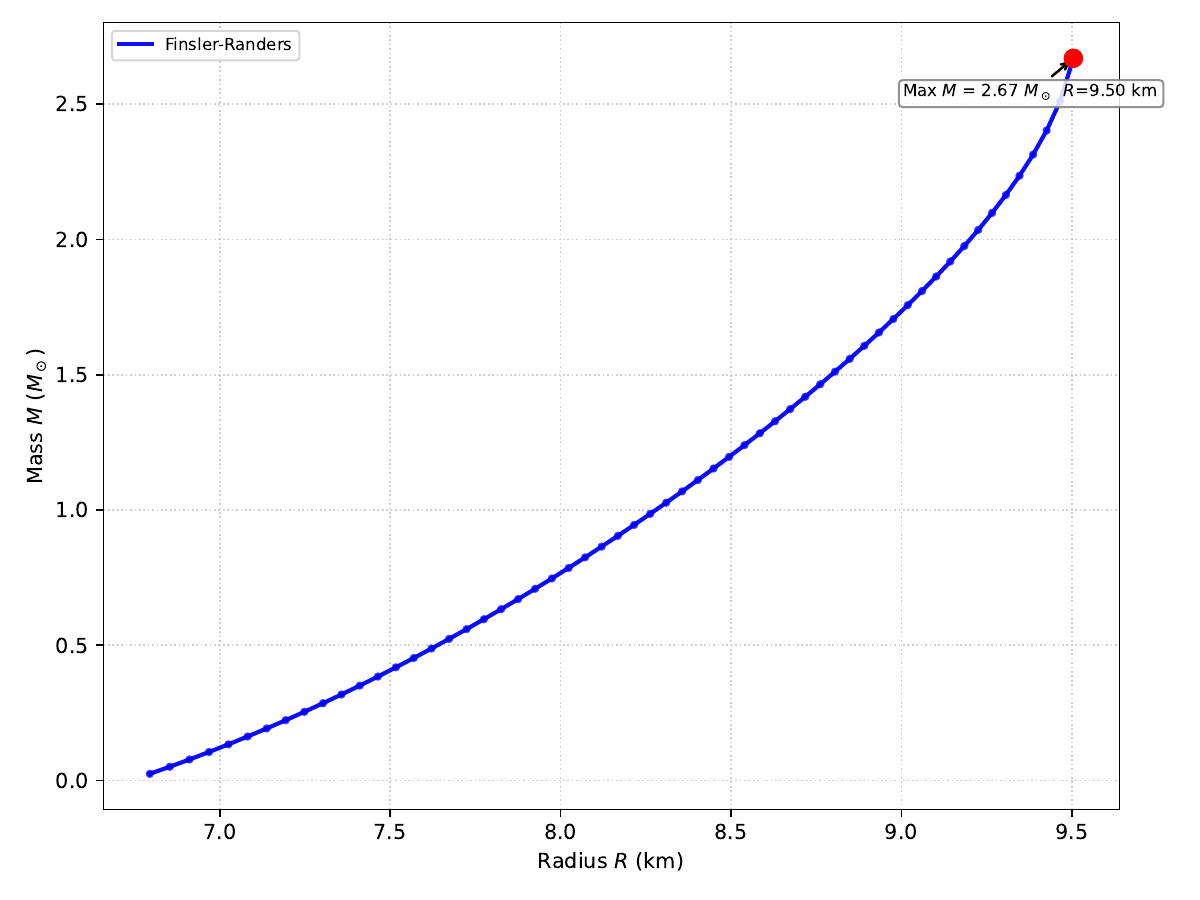} \\
(a) GR & (b) FR  \\
\end{tabular}
\caption{(a) M-R curve in GR framework (when $\xi =0$). The configuration attains a maximum mass of $M_{\text{max}}^{(\mathrm{GR})}=2.01\,M_{\odot}$ corresponding to a radius $R^{(\mathrm{GR})}=10.99\,\mathrm{km}$. (b) M-R curve in FR model. The configuration attains a maximum mass of $M_{\text{max}}^{(\mathrm{FR})}=2.67\,M_{\odot}$ at a radius $R^{(\mathrm{FR})}=9.50\,\mathrm{km}$}
\label{fig12}
\end{figure*}

Figures~\ref{fig12}(a) and (b) illustrate the $M–R$ profiles for the GR and FR configurations, respectively. In the GR case, the stellar mass increases with radius, attaining a maximum value of
\[
M_{\max}^{(\mathrm{GR})}=2.01\,M_{\odot}, \quad \text{at} \quad  R^{(\mathrm{GR})}=10.99\,\mathrm{km}.
\]
For our FR model exhibits a distinct sequence, with the mass reaching up to
\[
M_{\max}^{(\mathrm{FR})}=2.67\,M_{\odot}, \quad \text{at} \qquad R^{(\mathrm{FR})}=9.50\,\mathrm{km}.
\]
thereby allowing equilibrium configurations with significantly larger masses compared to the standard GR scenario.
\par
The observed enhancement in the maximum mass indicates that anisotropic corrections provide additional support against gravitational compression, enabling the system to maintain equilibrium at higher matter densities. In this context, the geometric anisotropy effectively contributes as an additional stabilizing factor in the hydrostatic balance, without necessitating any modification of the underlying EoS. This leads to an extension of the admissible stellar sequence toward more massive configurations.
\par
An equally notable outcome is that the FR configurations are not only more massive but also more compactness than their GR counterparts. From the $M–R$ curve profiles, the compactness increases from approximately
\[
\left(\frac{M}{R}\right)_{\mathrm{GR}}  \approx 0.183
\]
to
\[
\left(\frac{M}{R}\right)_{\mathrm{FR}}  \approx 0.281.
\]
This substantial increase demonstrates that the Finslerian corrections enhance the effective resistance to gravitational collapse, thereby supporting denser and more compactness stellar structures. A direct comparison of these two profiles clearly shows that the inclusion of FR geometric effects leads to significant modifications in the stellar configuration.
\par
From a physical perspective, the standard relativistic sequence reaches its limiting mass near $2\,M_{\odot}$, while the FR framework extends this bound appreciably ($2.67\,M_{\odot}$), thereby accommodating heavier stellar objects within the same matter composition. The gravitational-wave event \textit{GW190814}, detected by the LIGO/Virgo Collaboration, revealed a compact object with mass ($2.5-2.67\,M_{\odot}$)~\cite{abbott2020gw190814}, intensifying the debate on the existence of a maximum mass of neutron stars. Notably, the maximum mass predicted by our model falls within this observational range~\cite{abbott2020gw190814, tan2020neutron, wu2020if, godzieba2021maximum}, indicating that the FR framework can support ultra-massive compact stars without invoking exotic matter. Thus, the present model provides a plausible theoretical explanation for the nature of the massive compact objects observed in \textit{GW190814}, highlighting the role of geometric modifications of spacetime in explaining high-mass merger remnants.
\subsection{$I-M$ Relation: Comparative analysis with GR and its Astrophysical Relevance}
The Moment of Inertia-Mass (I-M) curve provides an additional and insightful probe into the internal structure and rotational properties of compact stars. In contrast to the mass–radius relation, which primarily characterizes global equilibrium, the moment of inertia is particularly sensitive to the radial distribution of matter within the stellar interior. As such, it serves as an effective diagnostic for assessing how different gravitational frameworks influence the internal mass profile and rotational behavior of neutron stars~\cite{ray1984rotating, malone1975neutron, baym1971ground}. For slowly rotating configurations, the moment of inertia $I$ can be reliably estimated using the approximate relation proposed by Bejger and Haensel~\cite{bejger2002moments},
\begin{equation}
    I = \frac{2}{5}(1+Y)MR^2
\end{equation}
which remains sufficiently accurate for slowly rotating compact objects, where parameter $Y = M/R. km/M_{\odot} $.
\par
\begin{figure*}[]
\centering
\setlength{\tabcolsep}{6pt}
\renewcommand{\arraystretch}{1.0}
\begin{tabular}{cc}
\includegraphics[width=0.48\textwidth]{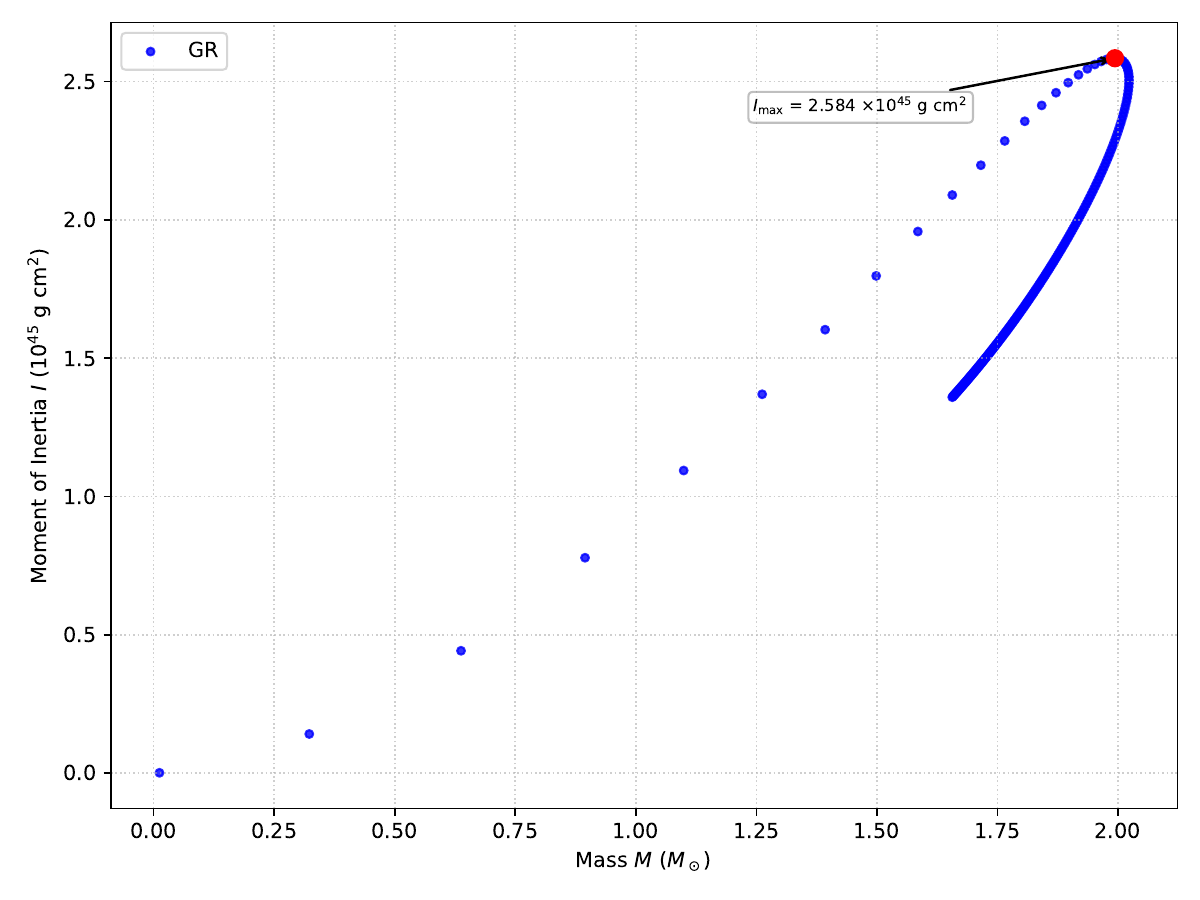} &
\includegraphics[width=0.48\textwidth]{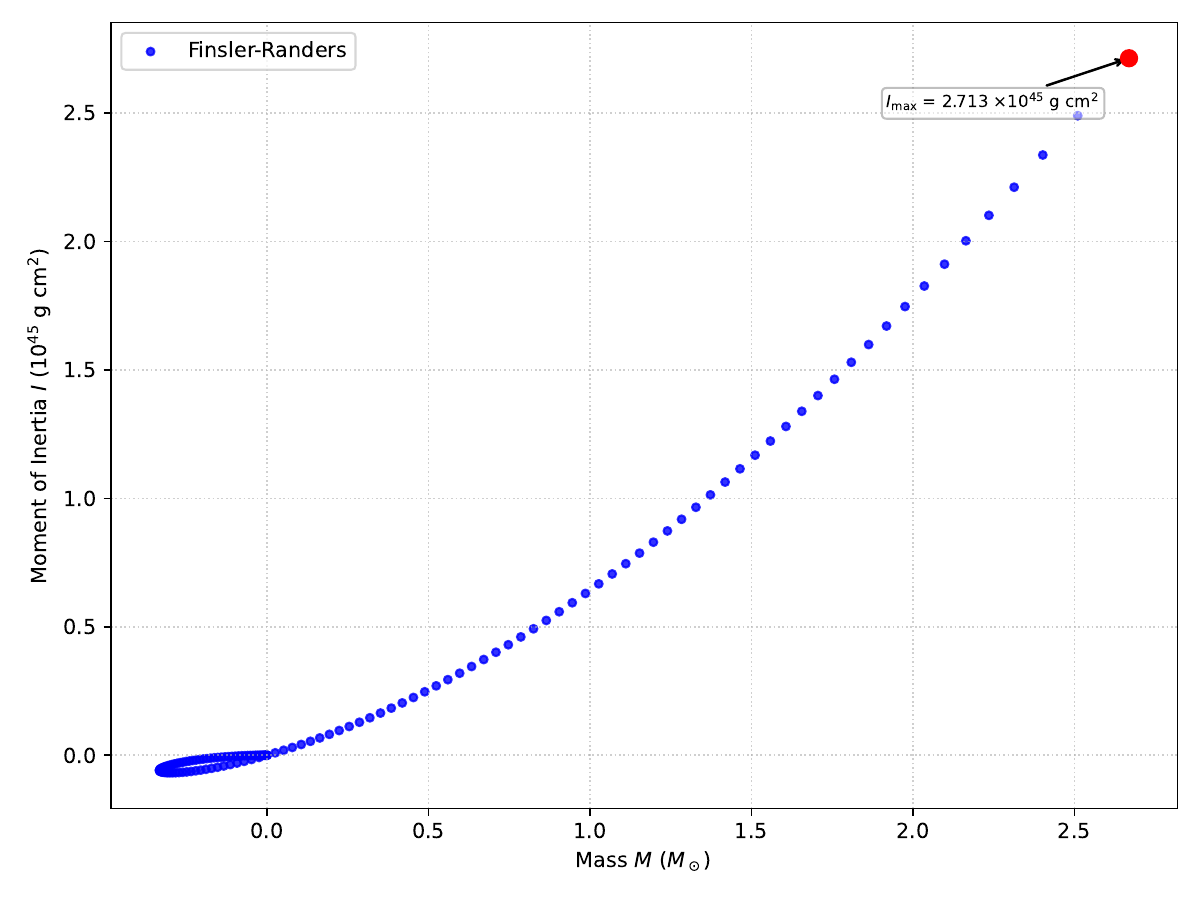} \\
(a) GR & (b) FR \\
\end{tabular}
\caption{(a) I-M curve in GR framework (when $\xi=0$). The configuration attains a maximum value of $I_{\text{max}} ^ {(\mathrm{GR})} = 2.53\times10^{45}\ , \mathrm{g\,cm^2}$, (b) I-M curve in the FR model. The configuration attains a higher maximum value of $I_{\text{max}}^ {(\mathrm{FR})}=2.71\times10^{45} \, \mathrm{g\,cm^2}$.}
\label{fig13}
\end{figure*}

Figures~\ref{fig13}(a) and (b) depict the variation of the moment of inertia with stellar mass for the GR and FR models, respectively. In the GR framework, the moment of inertia increases with mass and attains a maximum value of
\[
I_{\text{max}}^{(\mathrm{GR})} = 2.53 \times 10^{45}\,\mathrm{g\,cm^2}.
\]
On the other hand, the FR model exhibits a similar monotonic trend but reaches a higher maximum value of 
\[
I_{\text{max}}^{(\mathrm{FR})} = 2.71 \times 10^{45}\,\mathrm{g\,cm^2}.
\]
A direct comparison reveals that, for a given mass, the FR configurations consistently yield larger values of $I_{max}$, indicating an enhanced rotational nature relative to the GR case~\cite{zhang2020gw190814}. Rapidly rotating configurations have been proposed as a viable explanation for the secondary component of GW190814, suggesting that a pulsar spinning near its Keplerian (mass-shedding) limit could sustain masses in the range $2.5–2.67 M_\odot$ ~ \cite{zhang2020gw190814, breu2016maximum, wei2017rotating, most2020lower}. Physically, since the moment of inertia scales as $I \sim \int_0^R r^2\,dm$, even a modest outward redistribution of mass leads to a significant increase in rotational nature. Therefore, the larger values of $I$ obtained in the FR framework can be attributed to the effects of geometric anisotropy, which effectively promote a more extended mass distribution within the stellar interior.
\section{Boundary and Central Isotropy conditions: Model Parameters constraints from observed pulsars Data}\label{sec4}
\subsection{Boundary conditions}
The construction of realistic relativistic compact star models necessitates a smooth and consistent matching between the interior spacetime, which describes the matter distribution, and an appropriate exterior vacuum geometry. For static and spherically symmetric configurations, the exterior region is uniquely described by the Schwarzschild solution, which represents the only vacuum solution of the Einstein field equations under these symmetries. Consequently, it serves as the natural choice for modeling the exterior spacetime of compact objects such as neutron stars and strange stars~\cite{thorne2000gravitation, randers1941asymmetrical, bao2012introduction}.
\par
For such configurations, the exterior spacetime is therefore given by the Schwarzschild metric,
\begin{widetext}
\begin{equation}
ds^2_{+} = \left(1 - \frac{2M}{r}\right) dt^2 - \left(1 - \frac{2M}{r}\right)^{-1} dr^2 - r^{2} \left(d\theta^{2} + \sin^{2}\theta\, d\phi^{2}\right)
\label{eq37a}
\end{equation}
\end{widetext}
where $M$ represents the mass of the stellar system and here $G$, $c$ are chosen to be a unit.
\par
To ensure both physical and geometrical consistency at the stellar boundary $\Sigma: r = R $, the interior and exterior spacetimes must be smoothly joined. This matching is achieved through the continuity of the first and second fundamental forms, commonly referred to as the Darmois–Israel junction conditions~\cite{de1985collapse, santos1985non}. The continuity of the first fundamental form (the metric) ensures that the spacetime geometry remains smooth across the boundary, thereby avoiding coordinate or geometric singularities. Meanwhile, the continuity of the second fundamental form guarantees the absence of surface layers or thin shells, which would otherwise give rise to unphysical surface stresses.

At the boundary surface $\Sigma: r = R $, where the interior spacetime $ ds^2_{-} $ is matched to the exterior spacetime $ds^2_{+}$, the continuity of the first fundamental form requires that the metric components be continuous across $\Sigma $. In particular, the temporal and radial components of metric, must be continuous at the boundary, leading to the following boundary conditions:
\begin{equation}\label{eq37b}
    g_{tt}^{-} \overset{\Sigma}{=} g_{tt}^{+}  \implies e^{\mu(R)} = 1 - \frac{2M}{R}
    \end{equation}
    \begin{equation}\label{eq37c}
     g_{rr}^{-} \overset{\Sigma}{=} g_{rr}^{+} \implies e^{-\nu(R)} = 1 - \frac{2M}{R}
    \end{equation}
    \begin{equation}\label{eq37d}
   \frac{dg_{tt}^{-}}{dr} \overset{\Sigma}{=}  \frac{dg_{tt}^{+}}{dr} \implies \mu'(R) e^{\mu(R)} = \frac{2M}{R^2}
    \end{equation}
 where $r=R$ represents the radius of the compact star, $g_{ab}^{-}$ and $g_{ab}^{+}$ are the metric tensor components for interior (\ref{eq16a}) and exterior (\ref{eq37a}) spacetime respectively.
\par
The continuity of the second fundamental form yields the following condition~\cite{santos1985non}
    \begin{equation}\label{eq37e}
     p_r(R) =  0
    \end{equation}
This condition ensures that the radial pressure vanishes continuously at the stellar boundary, thereby establishing a consistent mechanical equilibrium between the interior FR configuration and the exterior Schwarzschild spacetime.

\subsection{Central isotropy condition}
In anisotropic stellar configurations, the radial pressure $p_r$ and tangential pressure $ p_t$ generally differ due to local anisotropies arising from dense matter interactions, phase transitions, or underlying geometric effects. Nevertheless, regularity at the stellar center imposes a fundamental constraint- the pressure must be isotropic at  $r = 0 $~\cite{herrera1997local}. Consequently, irrespective of the boundary conditions, the central region of a compact star must remain free from directional anisotropies.\\
The \textit{central isotropy condition} 
\begin{equation}
p_r(0) = p_t(0)
\label{eq37f}
\end{equation}
ensures the absence of directional dependence at the origin and requires that the anisotropy factor vanishes at $r = 0$~\cite{bowers1974anisotropic, dev2002anisotropic}. This condition follows naturally from spherical symmetry and the finiteness of all physical variables at the center, and is widely employed in anisotropic stellar models to fix integration constants or constrain model parameters.
\par
Equations~\eqref{eq37b}–\eqref{eq37e} and~\eqref{eq37f} collectively represent the necessary boundary and central isotropy conditions for the compact stellar configuration under consideration. These conditions play a crucial role in relativistic stellar modeling, as they allow for the determination of otherwise undetermined constants appearing in the model. By enforcing these constraints, the model parameters $(F, A, B, a, b)$ are uniquely calculated for the present configuration.
\par
By solving Eqs.~\eqref{eq37b}–\eqref{eq37f} simultaneously, one obtained as follows:
\begin{widetext}
\begin{align}
F &= \frac{\sqrt{3}\,(3R - 8M)}{R}\,\sqrt{\frac{M - R}{7M - 3R}}, \label{eq36}\\[6pt]
A &= \frac{1}{3\sqrt{3}}\,\sqrt{-\frac{(7M - 3R)^3}{R\,(R - 2M)^2}}, \label{eq37}\\[6pt]
B &= -\frac{M}{R^2\,(7M - 3R)}, \label{eq38}\\[6pt]
b &= -\frac{2a M\,(14M - 9R)}{3R^3\,(M - R)}, \label{eq39}\\[10pt]
a &= \frac{-72\sqrt{3}M^2\sqrt{\frac{M - R}{7M - 3R}}-27\sqrt{3}R^2\sqrt{\frac{M - R}{7M - 3R}}-27MR+99\sqrt{3}MR\sqrt{\frac{M - R}{7M - 3R}}+27R^2
}{216\sqrt{3}M^2\sqrt{\frac{M - R}{7M - 3R}}-392M^2+81\sqrt{3}R^2\sqrt{\frac{M - R}{7M - 3R}}-297\sqrt{3}MR\sqrt{\frac{M - R}{7M - 3R}}+393MR-81R^2} \label{eq40}
\end{align}
\end{widetext}
The foregoing relations uniquely determine the five parameters $(F, A, B, a, b)$ in terms of the stellar mass $M$ and radius $R$. In the present study, we employ observed pulsar data corresponding to the compact star candidates \textit{LMC X–4, Vela X–1, EXO 1785–248,} and \textit{Cen X–3 }~\cite{rawls2011refined, gangopadhyay2013strange, guver2010mass} to construct physically viable stellar configurations and to examine the influence of FR anisotropy on their internal structure. Utilizing the measured masses and radii, we have evaluated the numerical values of the parameters $(F, A, B, a, b)$ for each stars, as summarized in Table~\ref{Table1}. This procedure establishes a direct correspondence between the theoretical model and observational data, thereby enabling a consistent and realistic description of compact objects within the present framework. The resulting parameter sets, together with the imposed astrophysical constraints, are subsequently used to analyze key physical characteristics of the model.


\begin{table*}[]
\centering
\setlength{\tabcolsep}{10pt}
\renewcommand{\arraystretch}{1.2}
\caption{Estimated values of the model parameters $(A, B, F, b, a)$ corresponding to selected compact stars along with their observational masses $M$ and radii $R$}
\begin{tabular}{lccccccc}
\hline
\textbf{Star} & $M~(M_\odot)$ & $R~(\mathrm{km})$ & $A$ & $B$ & $F$ & $b$ & $a$ \\
\hline
LMC~X--4      & 1.40 & 8.30  & 0.794324 & 0.000856784 & 2.22092 & -0.00423698 & 0.421815 \\
Vela~X--1    & 1.77 & 9.55  & 0.679396 & 0.00119356  & 1.81784 & -0.00502924 & 0.472141 \\
EXO~1785--248    & 1.30 & 8.84  & 0.754188 & 0.000954972 & 2.07795 & -0.00446890 & 0.437715 \\
 Cen~X--3   & 1.49 & 9.178 & 0.725002 & 0.00103417  & 1.97553 & -0.00464684 & 0.450339 \\
\hline
\end{tabular}
\label{Table1}
\end{table*}

\section{Physical Viability of Finsler–Randers Stellar Model} \label{sec5}
\subsection{Energy Density and Pressure Profiles}\label{sec5.1}
\begin{figure*}[t]
\centering
\setlength{\tabcolsep}{6pt}
\renewcommand{\arraystretch}{1.0}
\begin{tabular}{cc}
\includegraphics[width=0.48\textwidth]{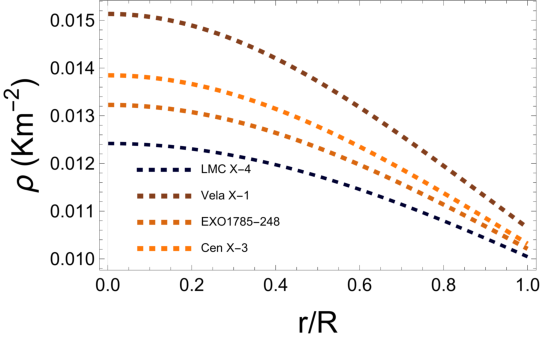} &
\includegraphics[width=0.48\textwidth]{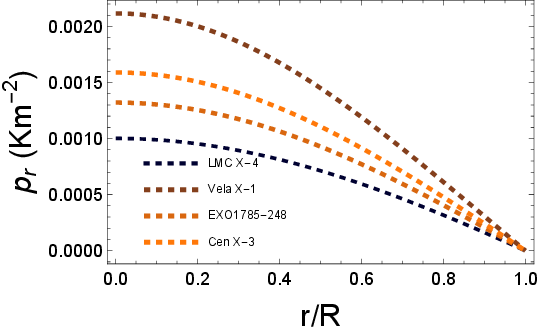} \\
(a) & (b)\\
\multicolumn{2}{c}{\includegraphics[width=0.52\textwidth]{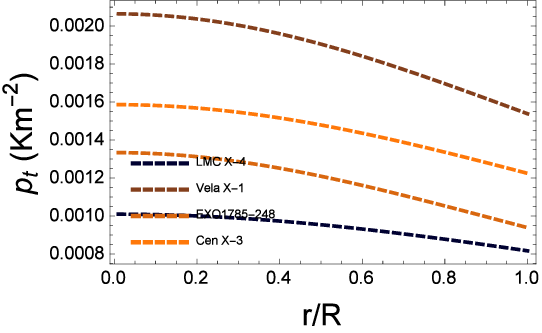}}\\
\multicolumn{2}{c}{(c)} \\
\end{tabular}
\caption{Variation of (a) energy density ($\rho$), (b) radial pressure ($p_r$), and (c) tangential pressure ($p_t$) for compact stars.}
\label{fig1}
\end{figure*}

Figure~\ref{fig1} illustrates the radial profiles of the energy density $\rho$, radial pressure $p_r$, and tangential pressure $p_t$ for the FR stellar model. All three quantities attain their maximum values at the stellar center and decrease monotonically toward the surface $r=R$, thereby indicating a well-behaved and physically admissible matter distribution. As evident from Fig.~\ref{fig1}(b,c), the model satisfies the necessary boundary condition, namely $ p_r(R)=0 $  in agreement with Eqs.~(\ref{eq37e}). Furthermore, the quantities $ \rho,  p_r $ and $p_t$ remain finite and positive throughout the stellar interior, ensuring regularity of the spacetime geometry and overall physical consistency within the FR framework. The corresponding central density $\rho_c $, surface density $\rho_s$, and central radial pressure $p_c$ are presented in Table~\ref{Table2}, providing a concise quantitative characterization of the internal structure of the compact configuration.
\subsection{Anisotropy Profile}
In relativistic stellar interiors, anisotropy arises when the radial and tangential pressures become unequal. Such deviations from isotropy are generally attributed to high-density effects, microscopic interactions, or modifications induced by the underlying geometric framework. The anisotropy parameter $\Delta$, defined by
\begin{equation}
\Delta = p_t - p_r
\label{eq38}
\end{equation}
\par
\begin{figure*}[]
\centering
\setlength{\tabcolsep}{6pt}
\renewcommand{\arraystretch}{1.0}
\includegraphics[width=0.5\textwidth]{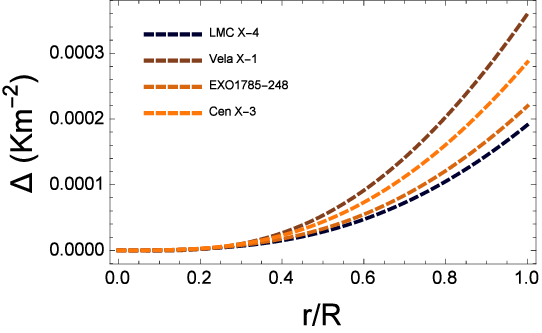} 
\caption{Profile of anisotropy $\Delta$.}
\label{fig2}
\end{figure*}

The radial variation of $\Delta $ is depicted in Fig.~\ref{fig2}. At the stellar center, the anisotropy vanishes, $\Delta(0)=0$, in accordance with the regularity and central-isotropy conditions. As the radial coordinate increases, $\Delta$ exhibits a monotonic growth and remains positive throughout the stellar interior. This behavior implies that the tangential pressure exceeds the radial pressure in the outer regions of the configuration. The resulting positive anisotropy introduces an effective repulsive force that counterbalances gravitational attraction, thereby providing additional support against collapse. Consequently, the system is able to sustain higher compactness and exhibits improved stability characteristics.

\subsection{Mass Function and Compactness}
The enclosed gravitational mass provides a direct measure of how matter is distributed within the stellar configuration and is defined as \cite{herrera1997local, buchdahl1959general}
\begin{equation}
m(r) = 4\pi \int_0^r \rho(x)\,x^2\,dx
\end{equation}
which represents the total energy enclosed within a radius $r$. The behavior of $m(r)$ therefore directly reflects how the density is distributed throughout the interior region.
\par
\begin{figure}[]
\centering
\includegraphics[width=0.5\textwidth]{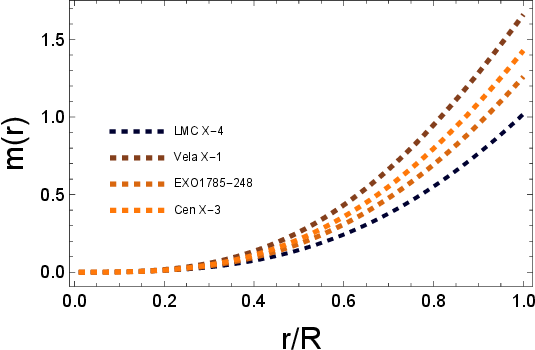}
\caption{Profile of gravitational mass function $m(r)$.}
\label{fig5}
\end{figure}
\begin{figure}[]
\centering
\includegraphics[width=0.5\textwidth]{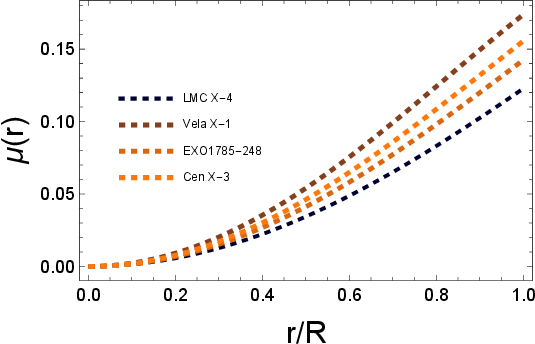}
\caption{Profile of compactness parameter $\mu(r)$. }
\label{fig6}
\end{figure}

The compactness parameter characterizes the degree of mass concentration within a given stellar radius and serves as a measure of the associated spacetime curvature. It provides a link between the internal mass distribution and the external gravitational field, thereby governing observable effects such as gravitational redshift and light deflection at the surface. For a static, spherically symmetric configuration, the compactness is defined as
\begin{equation}
\mu(r) = \frac{m(r)}{r}.
\end{equation}
As depicted in Figs.~\ref{fig5} and \ref{fig6}, both the mass function  $m(r)$ and the compactness $\mu(r)$ exhibit a monotonic increase from the center to the stellar surface, consistent with a positive and well-behaved energy density profile throughout the interior. For a physically viable relativistic configuration, the surface compactness must satisfy the Buchdahl bound, $\mu(R) < 4/9$ \cite{buchdahl1959general, andreasson2008sharp}, which imposes a fundamental upper limit on the mass–radius ratio of spherically symmetric matter distributions. The values of the surface-compactness $\mu_s$ obtained for the FR configuration, listed in Table~\ref{Table2}, are all found to satisfy this limit.

\par


\subsection{Gravitational Redshift}
\begin{figure}[]
\centering
\includegraphics[width=0.5\textwidth]{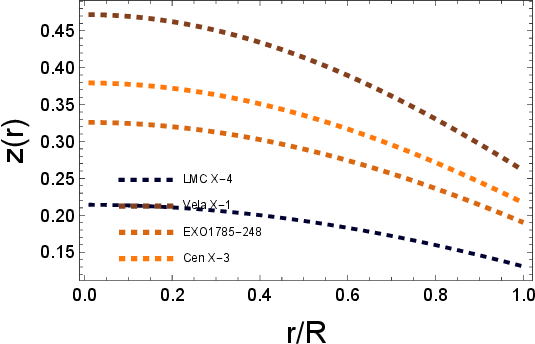}
\caption{ Profile of gravitational redshift $Z(r)$.}
\label{fig7}
\end{figure}

An important observable associated with compact objects is the gravitational redshift, which reflects the modification of photon frequencies as they propagate through the stellar interior. The redshift is given as,
\begin{equation}
Z(r) = \left[1 - 2\mu(r)\right]^{-1/2} - 1
\label{eq39}
\end{equation}
The variation of $Z(r)$ shown in Fig.~\ref{fig7} indicates that the gravitational redshift remains finite throughout the stellar interior, with a maximum at the center where the gravitational field is strongest, and a smooth decrease toward the surface. This behavior reflects the gradual reduction in spacetime curvature with radial distance. At the boundary, the surface redshift lies within the accepted theoretical bounds for compact stars, confirming the physical consistency of the model. The values of surface redshift $Z_s$ of the compact stars is listed in Table~\ref{Table2}.


\begin{table*}[]
\centering
\setlength{\tabcolsep}{8pt}   
\renewcommand{\arraystretch}{1.2} 
\caption{Estimated central density $\rho_c$, surface density $\rho_s$, central pressure $p_c$, adiabatic index $\Gamma_c$, surface compactness $\mu_s$, and surface redshift $z_s$ of the physical parameters of compact stars in Finsler--Randers (FR) model.}
\begin{tabular}{lcccccc}
\hline
Star & $\rho_c$ (g/cm$^3$) & $\rho_s$ (g/cm$^3$) & $p_c$ (dyne/cm$^2$) & $\Gamma_c$ & $\mu_s$ & $z_s$ \\
\hline
LMC~X--4  & $1.67 \times 10^{16}$ & $1.35 \times 10^{16}$ & $1.21 \times 10^{32}$ & $5.65$ & $1.2278 \times 10^{-1}$ & $1.5516 \times 10^{-1}$ \\
Vela~X--1 & $2.04 \times 10^{16}$ & $1.43 \times 10^{16}$ & $2.56 \times 10^{32}$ & $3.85$ & $1.7384 \times 10^{-1}$ & $2.6056 \times 10^{-1}$ \\
EXO~1785--248 & $1.78 \times 10^{16}$ & $1.37 \times 10^{16}$ & $1.60 \times 10^{32}$ & $4.82$ & $1.4248 \times 10^{-1}$ & $1.9024 \times 10^{-1}$ \\
Cen~X--3  & $1.86 \times 10^{16}$ & $1.39 \times 10^{16}$ & $1.92 \times 10^{32}$ & $4.37$ & $1.5562 \times 10^{-1}$ & $2.1688 \times 10^{-1}$ \\
\hline
\end{tabular}
\label{Table2}
\end{table*}

\section{Stability and Equilibrium Analysis of the model}\label{sec6}
\subsection{Energy Conditions}
The physical consistency of the matter distribution is further assessed through the verification of the standard energy conditions for anisotropic fluids. These conditions impose essential constraints on the relationship between the energy density and the principal pressures, thereby ensuring physically reasonable and stable behavior throughout the stellar interior.
\par
\begin{figure*}[]
\centering
\setlength{\tabcolsep}{8pt}
\renewcommand{\arraystretch}{1.0}
\begin{tabular}{cc}
\includegraphics[width=0.45\textwidth]{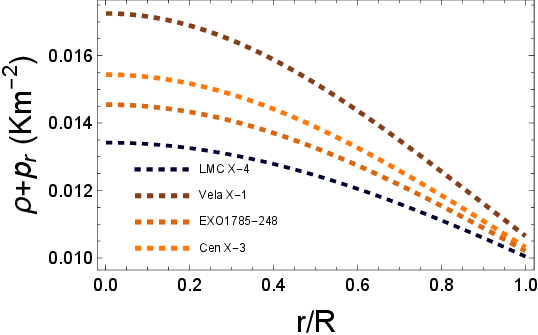} &
\includegraphics[width=0.45\textwidth]{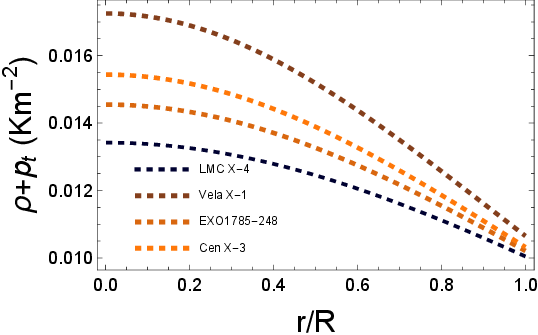} \\
(a) & (b)  \\[6pt]
\includegraphics[width=0.45\textwidth]{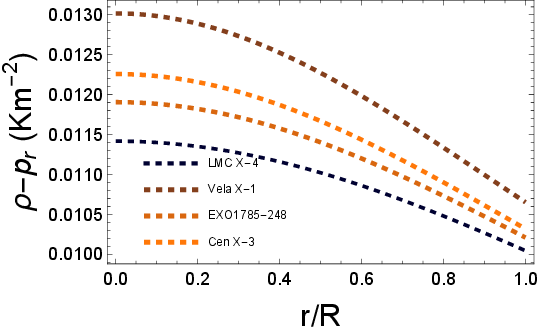} &
\includegraphics[width=0.45\textwidth]{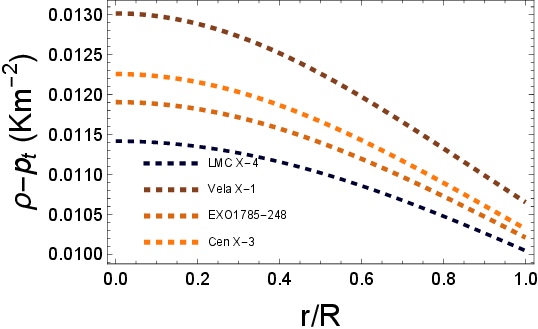} \\
(c)  & (d) \\[6pt]
\multicolumn{2}{c}{
\includegraphics[width=0.45\textwidth]{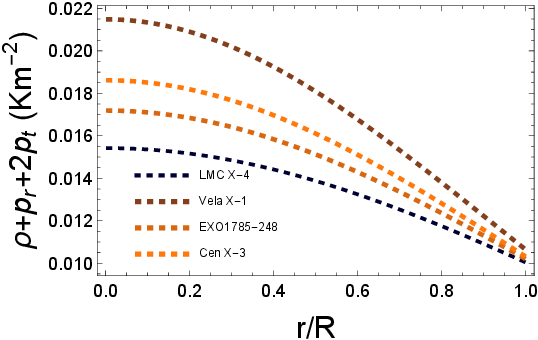}} \\
\multicolumn{2}{c}{(e)}
\end{tabular}
\caption{Energy Conditions.}
\label{fig3}
\end{figure*}

For the present configuration, these requirements are expressed in terms of the following inequalities:
\begin{align}
\text{NEC:} \quad & \{\rho + p_r,\; \rho + p_t\} \ge 0, \nonumber\\
\text{WEC:} \quad & \{\rho,\; \rho + p_r,\; \rho + p_t\} \ge 0, \nonumber\\
\text{SEC:} \quad & \{\rho + p_r,\; \rho + p_r + 2p_t\} \ge 0, \nonumber\\
\text{DEC:} \quad & \{\rho,\; \rho - p_r,\; \rho - p_t\} \ge 0.
\end{align}
The profiles displayed in Fig.~\ref{fig3} demonstrate that all energy conditions are satisfied across the entire radial domain, with the relevant combinations of energy density and pressures remaining positive throughout the interior. This behavior indicates that the matter distribution is physically viable and free from exotic contributions. Consequently, the model, including its Finslerian extension, remains well-behaved and consistent with fundamental physical requirements, without violating causality.
\subsection{Equation of State Parameter}
\begin{figure*}[]
\centering
\setlength{\tabcolsep}{6pt}
\renewcommand{\arraystretch}{1.0}
\begin{tabular}{cc}
\includegraphics[width=0.48\textwidth]{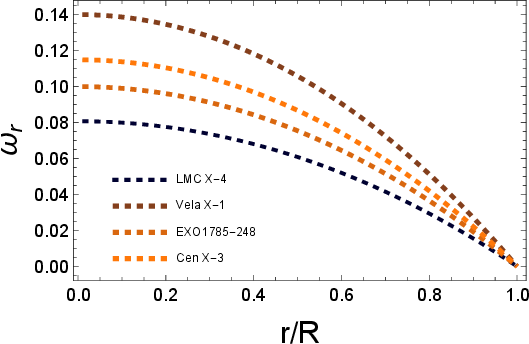} &
\includegraphics[width=0.48\textwidth]{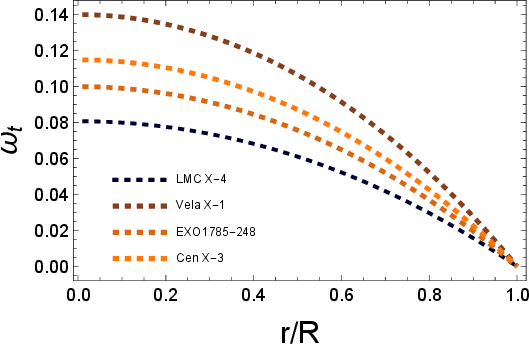} \\
(a) & (b) \\
\end{tabular}
\caption{Profiles of EoS parameters (a)  $\omega_r$  (b)  $\omega_t$.}
\label{fig4}
\end{figure*}

The physical nature of the fluid distribution can be examined through the equation of state (EoS) parameters, defined as 
\begin{align}
\omega_r = \frac{p_r}{\rho}, \qquad \omega_t = \frac{p_t}{\rho}.
\end{align}
The EoS parameters $\omega_r$ and $\omega_t$ provide important insights into the physical properties and composition of the stellar matter. For a physically realistic configuration composed of ordinary matter, these parameters must satisfy the condition $ 0 < \omega_r, \omega_t < 1$ throughout the stellar interior. As shown in Fig.~\ref{fig4}, both the radial and tangential EoS parameters remain well within this admissible range across the entire configuration. Although a deviation between  $\omega_r$ and $\omega_t$ emerges away from the center, the condition $\omega_r(0) = \omega_t(0)$ ensures regularity and isotropy at the stellar core. This divergence becomes more pronounced toward the boundary, reflecting the development of anisotropy in the outer layers. The smooth and well-behaved profiles of the EoS parameters thus affirm the physical consistency of the matter distribution within the FR-framework.

\subsection{ Sound's speed and Herrera’s cracking criterion}
The propagation of perturbations in an anisotropic stellar medium is governed by the corresponding sound speeds. The stability of such configurations can further be examined through the occurrence of cracking (or overturning) within the fluid distribution. Following the criterion proposed by Luis Herrera~\cite{herrera1992cracking}, local stability is determined by the response of anisotropic matter under perturbations, requiring the radial and tangential sound speeds to satisfy the causality conditions $0 < v_r^2 < c^2$ and $0 < v_t^2 < c^2$, where
\begin{equation}
v_r^2 = \frac{dp_r}{d\rho}, \qquad v_t^2 = \frac{dp_t}{d\rho}
\end{equation}
denote the radial and tangential sound velocities, respectively, and $c$ is the speed of light.
\par
\begin{figure*}
\centering
\setlength{\tabcolsep}{8pt}
\renewcommand{\arraystretch}{1.0}
\begin{tabular}{cc}
\includegraphics[width=0.45\textwidth]{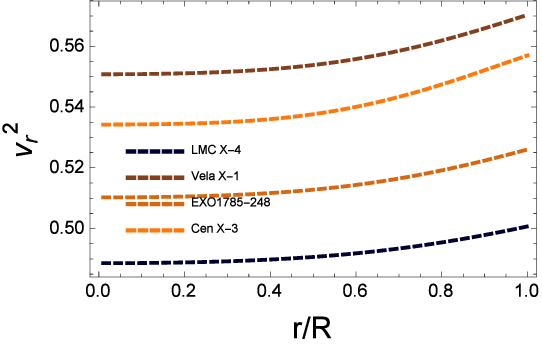} &
\includegraphics[width=0.45\textwidth]{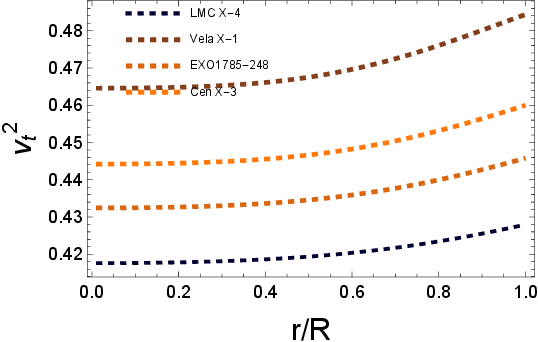} \\
(a) & (b) \\[6pt]
\multicolumn{2}{c}{%
\includegraphics[width=0.45\textwidth]{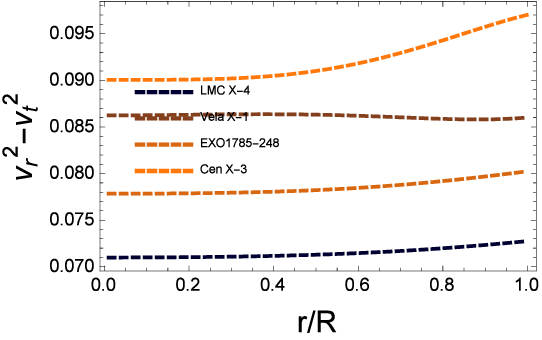}} \\
\multicolumn{2}{c}{(c)}
\end{tabular}
\caption{(a) radial sound speed ($v_r^2$), (b) tangential sound speed ($v_t^2$), and (c) $v_r^2 - v_t^2$: Herrera’s cracking criterion}
\label{fig9}
\end{figure*}

To evaluate the stability of the FR stellar model, we first impose the causality constraints $0 < v_r^2 < 1$ and $0 < v_t^2 < 1$ in relativistic units $(c = 1)$. As shown in Figs.~\ref{fig9}(a) and \ref{fig9}(b), both sound speeds remain subluminal throughout the stellar interior, thereby satisfying the causality condition. Furthermore, the Herrera’s cracking criterion $0 < v_r^2 - v_t^2 < 1$, which examines the response of the system to local perturbations, is satisfied and illustrated in Fig.~\ref{fig9}(c).

\subsection{Equilibrium analysis under the Tolman–Oppenheimer–Volkoff equation}
A physically viable stellar configuration must satisfy well-defined conditions that ensure equilibrium under the combined influence of gravitational forces and internal stresses. In this context, a systematic analysis of equilibrium is crucial for assessing the physical acceptability of anisotropic compact star models and for elucidating the role of anisotropy in maintaining hydrostatic balance in the presence of strong gravitational fields. The equilibrium of an anisotropic stellar system is governed by the relativistic hydrostatic condition, commonly formulated through the Tolman–Oppenheimer–Volkoff (TOV) equation \cite{tolman1939static, oppenheimer1939massive, mak2002exact, sharma2007class}.
\par
\begin{figure*}[]
\centering
\setlength{\tabcolsep}{8pt}
\renewcommand{\arraystretch}{1.0}
\begin{tabular}{cc}
\includegraphics[width=0.45\textwidth]{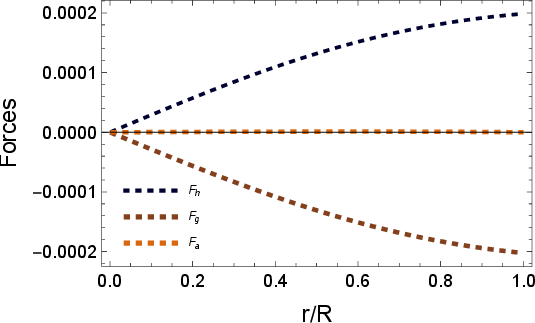} &
\includegraphics[width=0.45\textwidth]{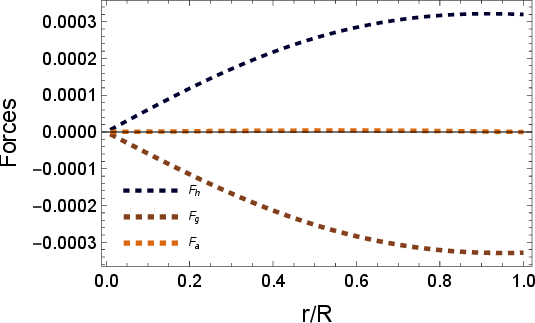} \\[-2pt]
(a)LMC~X--4 & (b) Vela~X--1 \\[6pt]
\includegraphics[width=0.45\textwidth]{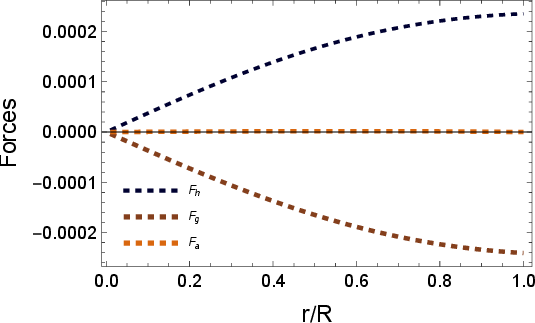} &
\includegraphics[width=0.45\textwidth]{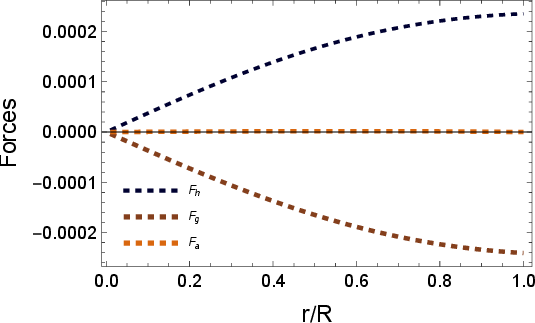} \\[-2pt]
(c) EXO~1785--248 & (d) Cen X--3 
\end{tabular}
\caption{Equilibrium condition: Graphical representation of the gravitational, anisotropic, and hydrostatic forces $(F_g, F_g, F_h)$.}
\label{fig8}
\end{figure*}

The TOV equation is expressed as follows:
\begin{equation}
-\frac{dp_r}{dr} - \frac{\mu'}{2}(\rho + p_r) + \frac{2}{r}(p_t - p_r) = 0,
\end{equation}
which encapsulates the interplay between pressure gradients, gravitational attraction, and anisotropic stresses. The above TOV equation may be rewritten in terms of three individual force components as:
\begin{align}
F_h &= -\frac{dp_r}{dr}, \\
F_g &= -\frac{\nu'}{2}(\rho + p_r), \\
F_a &= \frac{2}{r}(p_t - p_r),
\end{align}
where $F_h$, $F_g$, and $F_a$ represent the hydrostatic, gravitational, and anisotropic forces, respectively, which collectively satisfy the equilibrium condition
\begin{equation}
F_h + F_g + F_a = 0.
\end{equation}
Figure~\ref{fig8} demonstrates that the gravitational force remains negative throughout the stellar interior, indicating its inward-directed nature, while both the hydrostatic and anisotropic forces act outward. These competing forces effectively balance each other at every radial point, resulting in a vanishing net force. The mutual interplay among the gravitational, hydrostatic, and anisotropic forces $( F_g, F_h, F_a )$ thus ensures the maintenance of hydrostatic equilibrium across the entire configuration. This equilibrium condition substantiates the mechanical stability and supports the physical viability of the compact star model within the Finslerian framework.

\subsection{Adiabatic Index Analysis}
The adiabatic index serves as a fundamental diagnostic for assessing the dynamical stability of compact stars against infinitesimal radial perturbations~\cite{chandrasekhar1974slowly}. This criterion plays a central role in both Newtonian and relativistic descriptions of self-gravitating fluids. In the Newtonian limit, stability requires that the mean adiabatic index exceeds the critical value $\frac{4}{3}$  ~\cite{knutsen1988stability}. This condition was subsequently generalized to anisotropic matter distributions by Herrera et al.~\cite{herrera1992cracking, di1994tidal}, providing a more comprehensive framework for analyzing relativistic stellar models with directional pressures.
\par
\begin{figure}[]
\centering
\includegraphics[width=0.5\textwidth]{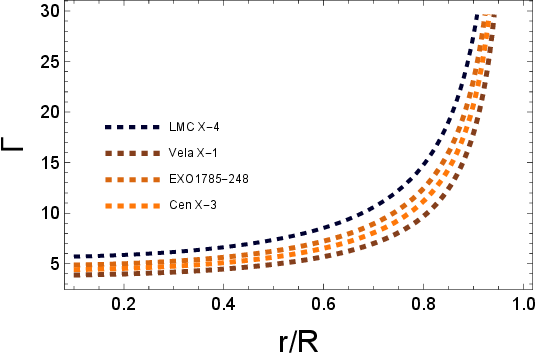}
\caption{ Profile of adiabatic index $\Gamma$. }
\label{fig10}
\end{figure}

The adiabatic index $\Gamma$ is defined as
\begin{equation}
\Gamma = \left( \frac{p_r + \rho}{p_r} \right) \frac{dp_r}{d\rho}.
\end{equation}
A relativistic stellar configuration remains dynamically stable when $\Gamma > \frac{4}{3}$, ensuring that pressure perturbations are sufficiently strong to counterbalance gravitational collapse~\cite{chandrasekhar1974slowly, heintzmann1975neutron}. For the present FR stellar model, the radial profile of $\Gamma$, depicted in Fig.~\ref{fig10}, exhibits a smooth and monotonically increasing trend from the center to the surface, remaining above the critical threshold throughout the interior. The corresponding central values of $\Gamma$ are listed in Table~\ref{Table2}, further emphasizing the stabilizing influence of Finslerian anisotropy on the stellar structure.

\subsection{Harrison--Zeldovich--Novikov criteria}
The stability of compact stellar configurations under radial perturbations is a fundamental requirement for their astrophysical viability. Among the standard diagnostics, the Harrison–Zeldovich–Novikov (HZN) criterion provides a simple yet physically transparent condition for assessing the stability of relativistic stars in hydrostatic equilibrium. When considered alongside other indicators—such as the adiabatic index, causality conditions on sound speeds, and force-balance analysis—it offers a complementary and robust measure of overall stability. The HZN criterion is based on the variation of the total gravitational mass (M) with respect to the central energy density $\rho_c$, requiring that a stable configuration satisfy $\frac{dM}{d\rho_c} > 0$~\cite{zeldovich1971relativistic, harrison1965gravitation}. A positive slope of the mass–central density relation implies that the system can respond to small perturbations without undergoing collapse, whereas a change in sign signals the onset of instability. Owing to its simplicity and direct connection with observable mass–radius characteristics, this criterion has become a widely used tool for validating compact star models, particularly in the presence of anisotropy or within modified gravitational frameworks~\cite{dev2003anisotropic, mak2003anisotropic}.
\par
\begin{figure*}[]
\centering
\begin{tabular}{cc}
\includegraphics[width=0.47\textwidth]{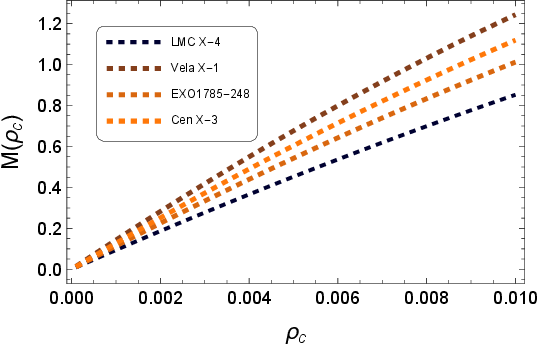} &
\includegraphics[width=0.47\textwidth]{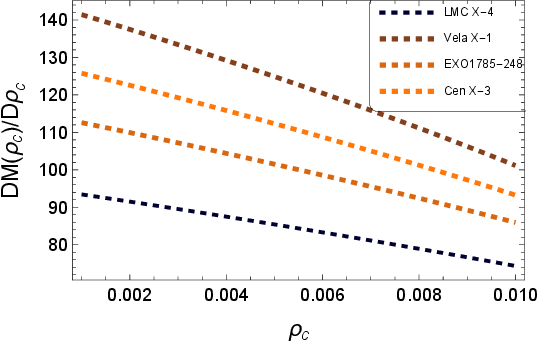}
\end{tabular}
\caption{Profiles of (a) mass–central density relation $M(\rho_c)$, and (b) $\frac{dM}{d\rho_c}$ with respect to central density $\rho_c$.}
\label{fig11}
\end{figure*}

The $M-\rho_c$ profiles profiles shown in Fig.~\ref{fig11}(a) exhibit a monotonic increase of mass with central density, indicating stable equilibrium throughout the considered range. This behavior is further corroborated in Fig.~\ref{fig11}(b), where the derivative $\frac{dM}{d\rho_c}$ remains strictly positive, thereby satisfying the HZN condition. Although a gradual decrease in the magnitude of the derivative is observed at higher central densities, it neither vanishes nor changes sign, ruling out the onset of instability.

\section{Conclusion and Summary of the work}\label{sec8}
The present analysis demonstrates that the FR gravity framework leads to a substantial modification of the global properties of compact stars, most notably in the prediction of the maximum mass, compactness, and moment of inertia. Subsequently, the Mass--Radius (M--R) relation is constructed to examine equilibrium configurations and identify the maximum mass limit.  For the class of solutions constructed using the Heintzmann IIa metric potentials and a linear equation of state, we obtain stable configurations with maximum mass reaching $ \sim 2.67 M_\odot$. This upper bound lies within the mass range inferred for the secondary component of \textit{GW190814} and is therefore consistent with current GW observations of high-mass compact objects. In addition, the model predicts an increase in the maximum compactness compared to the GR, indicating that more tightly bound configurations can be supported within the FR framework. This behavior reflects the effective contribution of spacetime anisotropy, which modifies the internal pressure gradients and allows equilibrium configurations at higher central densities without violating causality or stability conditions. Correspondingly, the moment of inertia is also found to increase, with the $I–M$ relation exhibiting systematically larger values than in GR. This suggests stronger rotational support and a nontrivial redistribution of the internal mass profile, which could have observable consequences for rapidly rotating pulsars.
\par
The compatibility of these results with GW observations is particularly noteworthy and the ability of the model to support maximum masses in the range $2.5-2.67 M_\odot$, without invoking exotic matter components or extreme modifications of the microphysical equation of state, provides a natural explanation for compact objects residing in the proposed mass gap. Instead, the required enhancement arises from the underlying Finslerian geometric structure, which effectively encodes additional degrees of freedom beyond standard Riemannian geometry. Furthermore, the constructed FR model satisfy all necessary physical requirements, including regularity, energy conditions (density, pressures, anisotropy), red-shift anlysis, subluminal sound speeds, and equilibrium under the TOV equation. Stability is confirmed through the EoS, adiabatic index, the Harrison–Zeldovich–Novikov criterion, and cracking analysis, ensuring that the obtained configurations are not only massive but also dynamically stable.
\par
These results indicate that FR gravity provides a viable extension of GR capable of reconciling theoretical predictions with current GW observations, and offers a promising avenue for probing spacetime geometry through high-precision measurements of compact star properties. The present work also opens several avenues for future investigation. The model can be extended by incorporating more realistic EoS derived from nuclear physics and by including effects of rapid rotation and magnetic fields. Further studies may also explore observational constraints using recent data from pulsars, GW, X-ray measurements, etc. to test the viability of Finslerian compact star models. In addition, this framework may be generalized to other astrophysical systems such as quark stars, hybrid stars, and anisotropic fluid spheres and such developments would deepen the understanding of the role of spacetime anisotropy in strong gravitational regimes.
\par
\textbf{Funding statement:}    No funding was received for this research.
\par
\textbf{Data Availability Statement:} Data associated in the manuscript will be available on request.
\par
\textbf{Conflict of interest:} The authors confirm that there are no financial interests or personal affiliations that could have influenced the research presented in this paper.
\par
\textbf{Ethics approval statement:} This work is original and has not been published elsewhere in any form or language (partially or in full). All authors read and approved the final manuscript.
\par
\textbf{Author Contribution: } Praveen J: Conceptualization, Formal analysis, Methodology, Writing-original draft. Rajesh Kumar: Conceptualization, Methodology, Investigation, Validation, Review and Editing. Sadaf Fatima: Conceptualization, Methodology, Investigation, Validation, Review and Editing.S K Narasimhamurthy: Visualization, Supervision, Project administration, Investigation.  
\par
\textbf{Acknowledgment:} The author  RK is thankful to IUCCA, Pune, India, for providing facilities under associateship programs where a part of the work is done during his the visit.

\bibliography{Referecnces}    
\end{document}